\documentclass{PoS}

\usepackage{amsmath,amssymb,amsfonts}
\usepackage[english]{babel}

\newcommand{\A}{\textrm{a}\,}

\newcommand{\be}{\begin{equation}}
\newcommand{\ee}{\end{equation}}
\newcommand{\bea}{\begin{eqnarray}} 
\newcommand{\eea}{\end{eqnarray}}

\newcommand{\bmp}{\noindent\begin{minipage}{16cm}}

\newcommand{\kari}{}

\title{The gradient flow running coupling in SU2 with 8 flavors}

\ShortTitle{The gradient flow running coupling in SU2 with 8 flavors}

\author{\speaker{Jarno Rantaharju}\\
       RIKEN Advanced Institute of Computational Science and\\
       CP3 -Origins, IFK \& IMADA, University of Southern Denmark\\
       E-mail: \email{rantaharju@cp3.dias.sdu.dk}}

\author{Tuomas Karavirta\\
        {CP}$^{ \bf 3}${-Origins} \& DIAS, University of Southern Denmark,\\ 
Campusvej 55, DK-5230 Odense M, Denmark.\\
        E-mail: \email{karavirta@cp3-origins.net}}

\author{Viljami Leino\\
        Helsinki Institute of Physics and Department of Physics University of Helsinki\\
        E-mail: \email{viljami.leino@helsinki.fi}}

\author{Teemu Rantalaiho\\
        Helsinki Institute of Physics and Department of Physics University of Helsinki\\
        E-mail: \email{teemu.rantalaiho@helsinki.fi}}

\author{Kari Rummukainen\\
        Helsinki Institute of Physics and Department of Physics University of Helsinki\\
        E-mail: \email{kari.rummukainen@helsinki.fi}}

\author{Kimmo Tuominen\\
        Helsinki Institute of Physics and Department of Physics University of Helsinki\\
        E-mail: \email{kimmo.i.tuominen@helsinki.fi}}

\abstract{We present preliminary results of the gradient flow running coupling with Dirichlet boundary condition in the SU(2) gauge theory with 8 fermion flavours. Improvements to the gradient flow measurement allow us to obtain a robust continuum limit. The results are consistent with perturbative running in the weak coupling region.
\vskip 0.2cm
{\noindent \footnotesize Preprint: CP3-Origins-2014-038, DIAS-2014-38}
}

\FullConference{The 32nd International Symposium on Lattice Field Theory\\
		 23-28 June, 2014\\
		 Columbia University New York, NY}

\begin{document}

\section{Introduction}

In the space of gauge theories there is a class of {\kari asymptotically free} models where the running coupling approaches {\kari a non-trivial fixed point} in the infrared {\kari and long-distance physics are conformal.}  In terms of the number of fermion flavours this class of models is limited from above by the loss of asymptotic freedom as the leading coefficient of the perturbative expansion of the $\beta$-function becomes positive. It is also limited from below by the onset of chiral symmetry breaking. These conformal models have applications in model building beyond the standard model, particularly in technicolor theories, where the electroweak symmetry is broken by  
{\kari the formation of the chiral condensate} in a strongly interacting model. They are also interesting from the theoretical point of view of understanding the structure of gauge field theories.

While perturbation theory gives an accurate description of the loss of asymptotic freedom, the lower limit depends on whether the chiral symmetry is broken before the fixed point is reached in the infrared. This {\kari typically} happens at large coupling, where perturbation theory is not expected to give an accurate description of the model.


In this proceedings we study the running of the coupling in SU(2) gauge field theory with 8 fermions in the fundamental representation. {\kari Based on perturbative analysis,} this model is expected to lie within the conformal window, but close to the lower limit. It was studied before in \cite{Ohki:2010sr} but the results were inconclusive. The phase structure of the model with staggered fermions was studied in these proceedings by {\kari Huang et al.} \cite{Huang:2014xwa}.

The running of the coupling can be studied directly on the lattice by measuring the coupling at different renormalisation scales, {\kari with the help of background fields generated by the Schrödinger functional} \cite{Luscher:1991wu}.  {\kari This method has been applied to SU(2) gauge theory with 4--10 fundamental representation fermion flavours \cite{Bursa:2010xn,Karavirta:2011zg,Hayakawa:2013maa,Appelquist:2013pqa}, but with inconclusive results regarding the location of the conformal window.  Because the running is slow in theories close to the conformal window,  very high accuracy is needed to discern the continuum behaviour, making simulations at large volumes prohibitive.}  

The gradient flow method {\kari \cite{Luscher:2011bx}
promises a definition of the coupling using a considerably less noisy observable.  In this work we use the gradient flow with Schrödinger functional boundary conditions \cite{Fritzsch:2013je}.  It also enables us to control the leading order discretisation errors and hence improve the continuum limit.}

\section{Methods and Results}

We study the model using a HEX smeared \cite{Capitani:2006ni}, clover improved Wilson fermion action and a partially smeared plaquette gauge action. The full action can be written as
 \begin{align*}
   S = (1-c_g)S_G(U) + c_g S_G(V) + S_F(V) + c_{SW} \delta S_{SW}(V),
  \end{align*}
where $V$ is the smeared gauge field and $U$ is the unsmeared one. As a result of the smearing, the action is non-perturbatively order $a$-improved when {\kari the Sheikholeslami-Wohlert coefficient} $c_{SW}\approx1$ and we simply choose $c_{SW}=1$. The gauge {\kari action} smearing, tuned by the coefficient $c_g$, removes the unphysical bulk phase transition from the region of interest in the parameters space. In this case it is sufficient to choose $c_g=0.5$.

We also use Schr\"odinger Functional boundary conditions, which enables us to measure the mass anomalous dimension of the model from the same dataset. The gauge fields are set to unity and the fermion fields are set to zero at {\kari time slices $x_0=0,L$ on a lattice of size $L^4$}:
\begin{align*}
  &U_k(x) = V_k(x) = 1, ~~~\psi(x) = 0  \text{~~ when ~~} x_0=0,L \\
  &U_\mu(x+L\hat k) = U_\mu(x), ~~~   V_\mu(x+L\hat k) = V_\mu(x)\\
  &\psi(x + L\hat k) = \psi(x).
\end{align*}
{\kari Here $k=1,2,3$ labels one of the spatial directions.}

The running coupling can be measured from the gradient flow evolution of the gauge field in
a fictitious flow time along the gradient of the action.
In continuum notation the flow is 
\begin{align*}
  \partial_t B_{t,\mu} &= D_{t,\mu} B_{t,\mu\nu}, \\ B_{0,\mu} &= A_\mu\\
  B_{t,\mu\nu} &= \partial_\mu B_{t,\nu} - \partial_\nu B_{t,\mu} 
  + \left[ B_{t,\mu},B_{t,\nu} \right].
\end{align*}
Here $B_{t,\mu}$ is the flow field parametrised by the flow time $t$, and $A_\mu$ is the original gauge field.
The flow smooths the field, moving toward the minimum of the action.
Correlators of the flow field are automatically renormalised
and therefore encode physical properties of the theory \cite{Luscher:2011bx}.

The coupling is measured from the evolution of the field strength
\begin{align*}
  \left<E(t)\right> &= \frac 14 \left<G_{\mu\nu}(t)G_{\mu\nu}(t)\right>.
\end{align*}
To the leading order in perturbation theory, it has the form \cite{Luscher:2010iy}
$
  \left<E(t)\right> = N g^2/t^2 + \mathcal{O}(g^4).
$
The observable can therefore be used to define a renormalised coupling \cite{Fodor:2012qh},
\begin{align} \label{gf_coupling}
  &g^2_{GF} = \frac{t^2 \left < E(t) \right>}{N}.
\end{align}

We study the running of the coupling with the renormalisation scale
by measuring the coupling with several physical lattice sizes.
To quantify the running we use the step scaling function \cite{Luscher:1993gh}
\begin{align}\label{lat_step}
  &\Sigma(u,\A/L) = \left . g_{GF}^2(g_0,2L/\A) \right|_{g_{GF}^2(g_0,L/\A)=u}\\\nonumber
  &\sigma(u) = \lim_{\A \rightarrow 0} \Sigma(u,\A/L). \nonumber
\end{align}
The step scaling function describes how the coupling evolves when
the linear size of the system is increased.
The gradient flow introduces another length scale, $l=\sqrt{8t}$,
which we fix to {\kari be proportional to the} lattice scale as $l=c_tL$.

In this study we use the Symanzik gauge action to generate the gradient flow on the lattice. The translation symmetry is broken by the boundary conditions and we hope to avoid any unnecessary finite size effects by measuring the coupling only at the middle time slice,
\begin{align*}
   N(c_t,a/L) g^2_{GF} = t^2 \left < E(t,x_0) \right>,  \,\,\,\,\, x_0=L/2, \,\, t = (c_t L)^2/8.
 \end{align*}
The normalisation  factor $N$ for this formulation has been calculated in \cite{Fritzsch:2013je}.
{\kari Unless otherwise indicated we use $c_t=0.4$ in our analysis below.}

\begin{figure}
  \center
  \includegraphics[width=0.45\textwidth]{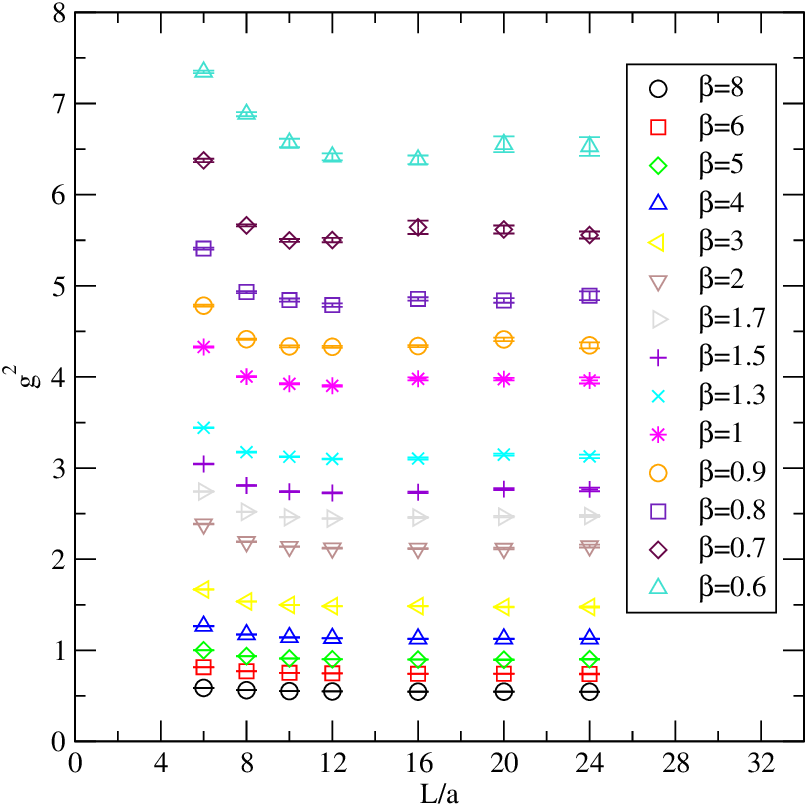} 
\caption[b]{
  The gradient flow coupling (\ref{gf_coupling}) at each $\beta$ and $L/a$
  {\kari at $c_t=0.4$}.
}
\label{fig:g2_lat_meas}
\end{figure}

\begin{figure}
  \includegraphics[width=0.45\textwidth]{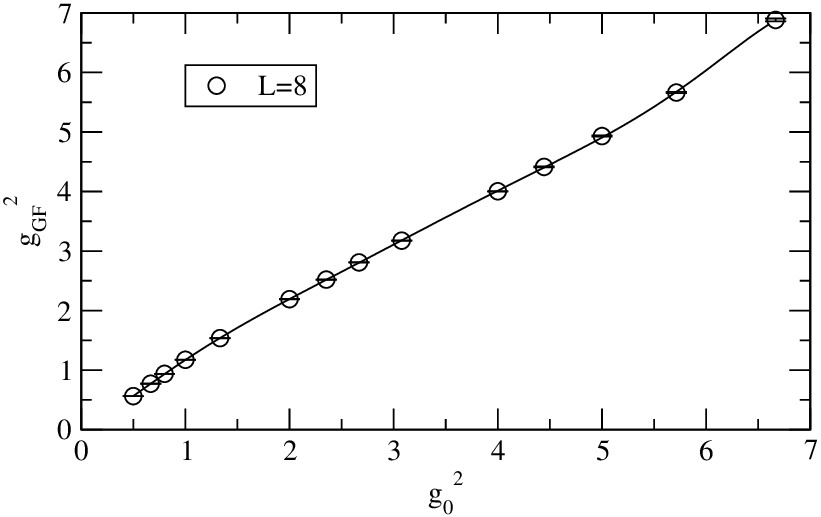} 
  \includegraphics[width=0.45\textwidth]{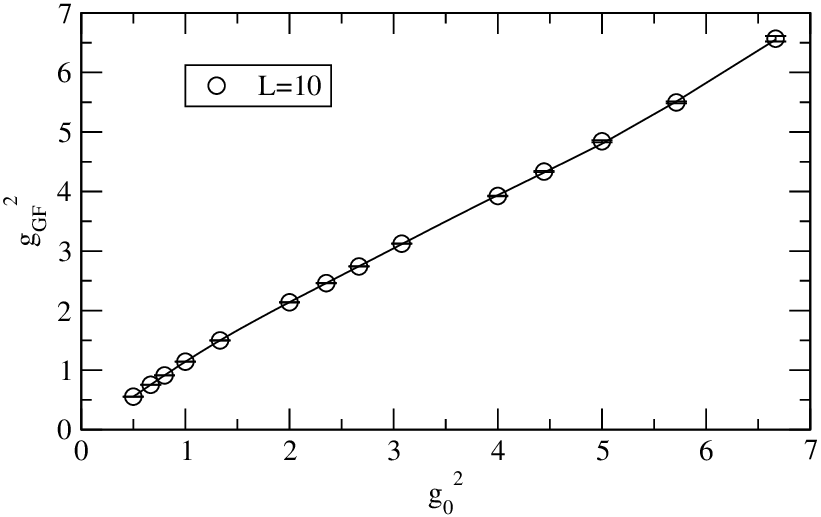}
\caption[b]{
	The gradient flow coupling and the interpolating function (\ref{betafitfun}).
  }
\label{fig:g2_interpolate}
\end{figure}

{\kari The measured values of the coupling are shown in figure~\ref{fig:g2_lat_meas}.}
In order to study the continuum limit of the step scaling function
we need to obtain it at constant values of the coupling
at several lattice sizes.
This is achieved with an interpolating function of the form
\begin{align} \label{betafitfun}
  g^2_{GF}&(g_0,\A/L) = g_0^2 \frac{ 1 + \sum_{i=1}^m
  a_i g_0^{2i}}{ 1 + \sum_{i=1}^n
  b_i g_0^{2i}} , \,\,\,\, m=4,n=3,
\end{align}
{\kari see figure~\ref{fig:g2_interpolate}.}
The continuum limit $\sigma(u)$ can then be approximated by fitting to a function of $a/L$,
\begin{align}\label{contfit}
  &\Sigma(u,\A/L)  = \sigma(u) + c_2(u) (a/L)^2.
\end{align}

\begin{figure}
  \center
  \includegraphics[width=0.45\textwidth]{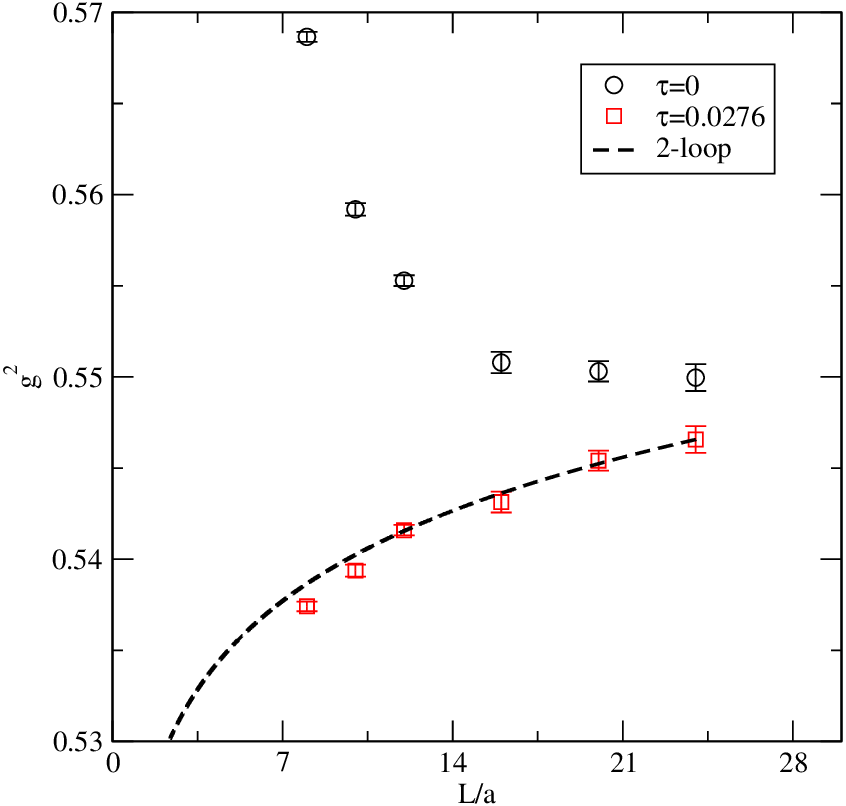} 
\caption[b]{
  The effect of the correction $\tau_0$ on the gradient flow coupling at $\beta=8$.
}
\label{fig:b8_tau0}
\end{figure}

\begin{figure}
  \includegraphics[width=0.45\textwidth]{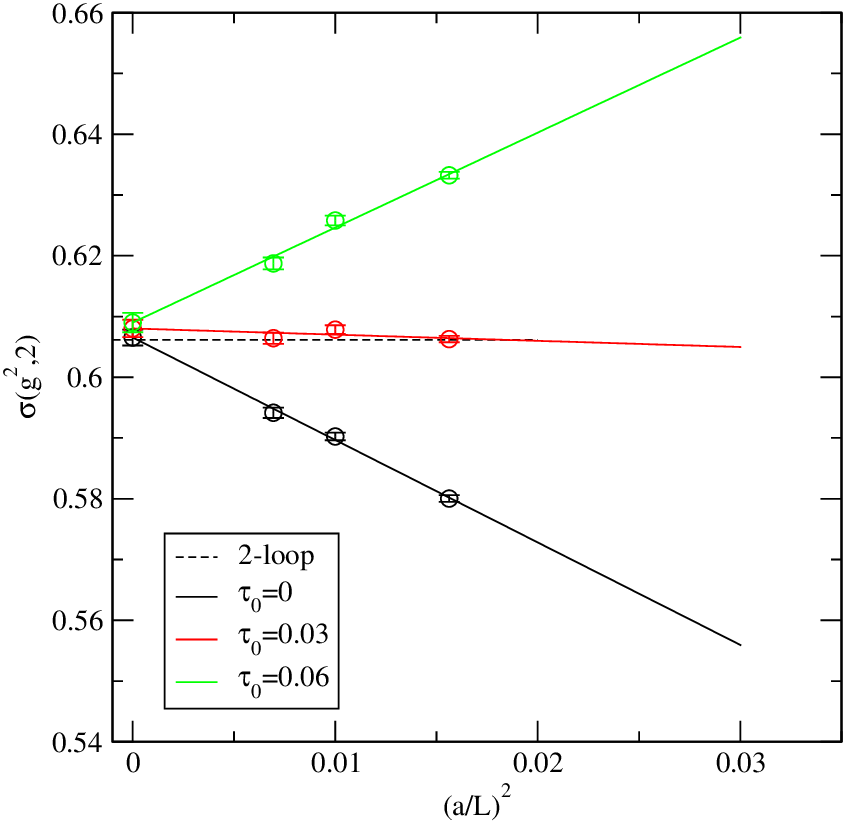} 
  \includegraphics[width=0.45\textwidth]{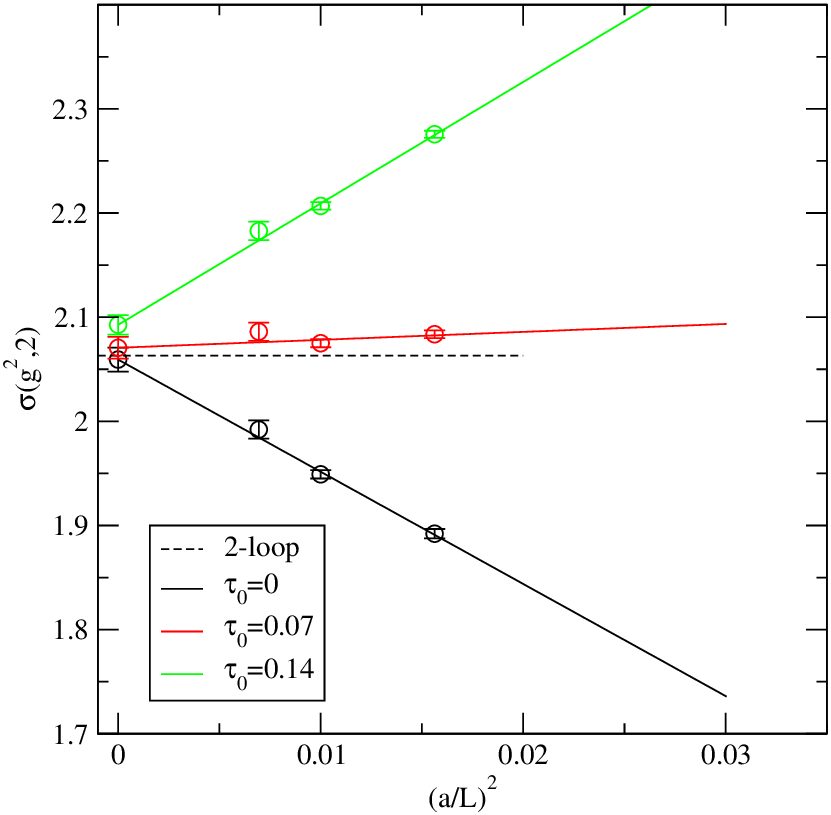}
\caption[b]{
  The continuum limit (\ref{contfit}) with several values of the flow time correction $\tau_0$
  with $g^2=0.6$ (left) and $g^2=2$ (right).
}
\label{fig:sigma_fit_tau}
\end{figure}

As was observed in\cite{Rantaharju:2013bva}, the gradient flow coupling suffers from large order $(a/L)^2$ errors. These can be alleviated with a correction $\tau_0$ to the lattice value of the flow time \cite{Cheng:2014jba}
\begin{align} \label{taucorrection}
      g^2_{GF} &=  \frac{t^2}{N} \left < E(t+\tau_0 a^2) \right>  
      = \frac{t^2}{N} \left < E(t) \right> 
      + \frac{t^2}{N} \left < \frac{\partial  E(t)  }{ \partial t} \right> {\tau_0} a^2.
\end{align}
The effect of the correction on the coupling and the step scaling function can be seen in figures \ref{fig:b8_tau0} and \ref{fig:sigma_fit_tau} respectively. As can be seen in figure \ref{fig:sigma_fit_tau}, the correction has a large effect on the $(a/L)^2$ component of the lattice step scaling function, but only a small one on any higher order components. As such, the exact value of the correction is not significant, and we can choose a value that is close enough in the measured range of couplings. In the following results we choose a functional form to approximate the correction. For example, for $c_t=0.4$ we use
\begin{align} \label{taufunc}
       \tau_0 = 0.064 \log (1+g^2).
\end{align}

\begin{figure}
  \includegraphics[width=0.45\textwidth]{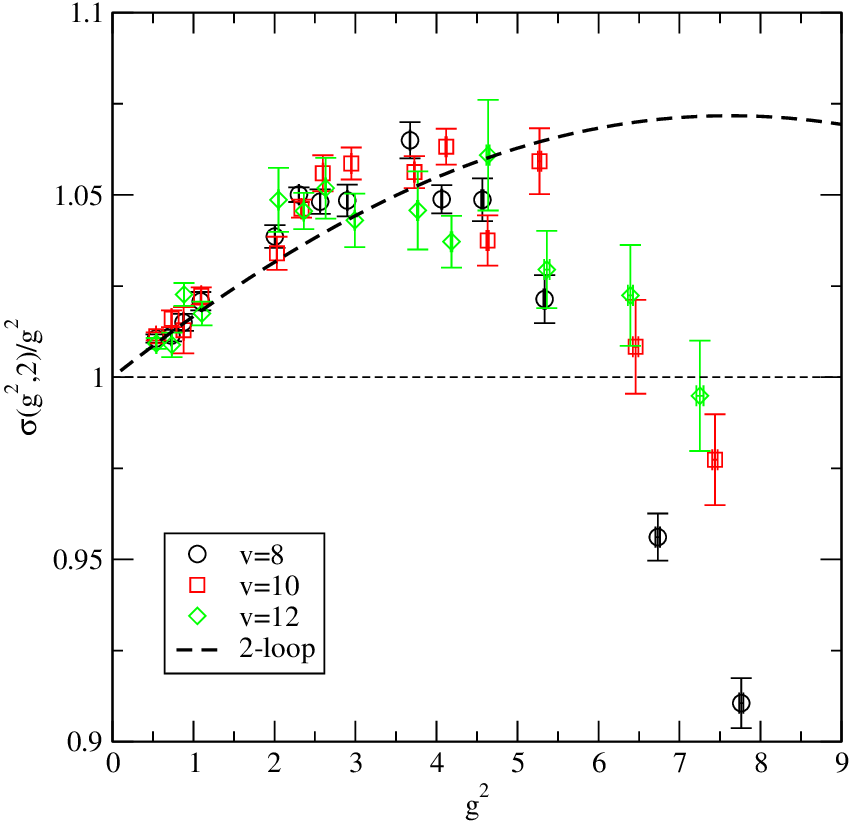} 
  \includegraphics[width=0.45\textwidth]{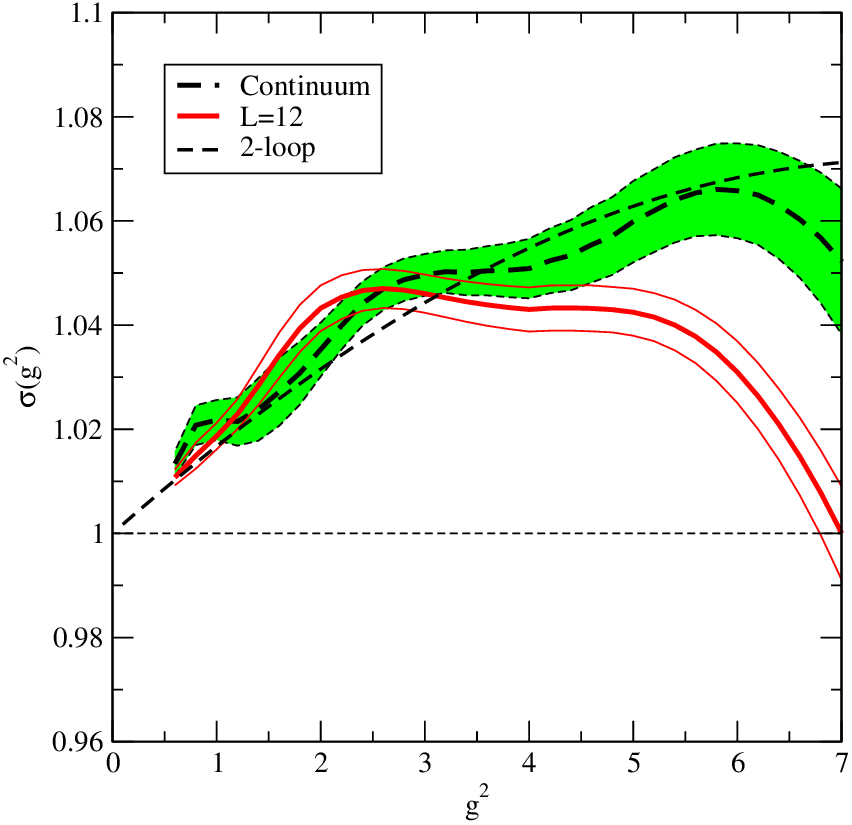}
\caption[b]{
  The step scaling function with $c_t=0.4, \tau_0 = 0.064\log(1+u) $.
  The plot on the left shows the lattice step scaling function (\ref{lat_step}) 
  and the one on the right shows the continuous step scaling function
  calculated from the fit (\ref{betafitfun})
  at the largest lattice size
  and the continuum limit (\ref{contfit}). 
}
\label{fig:sigma_lat_cont}
\end{figure}

\begin{figure}
  \includegraphics[width=0.45\textwidth]{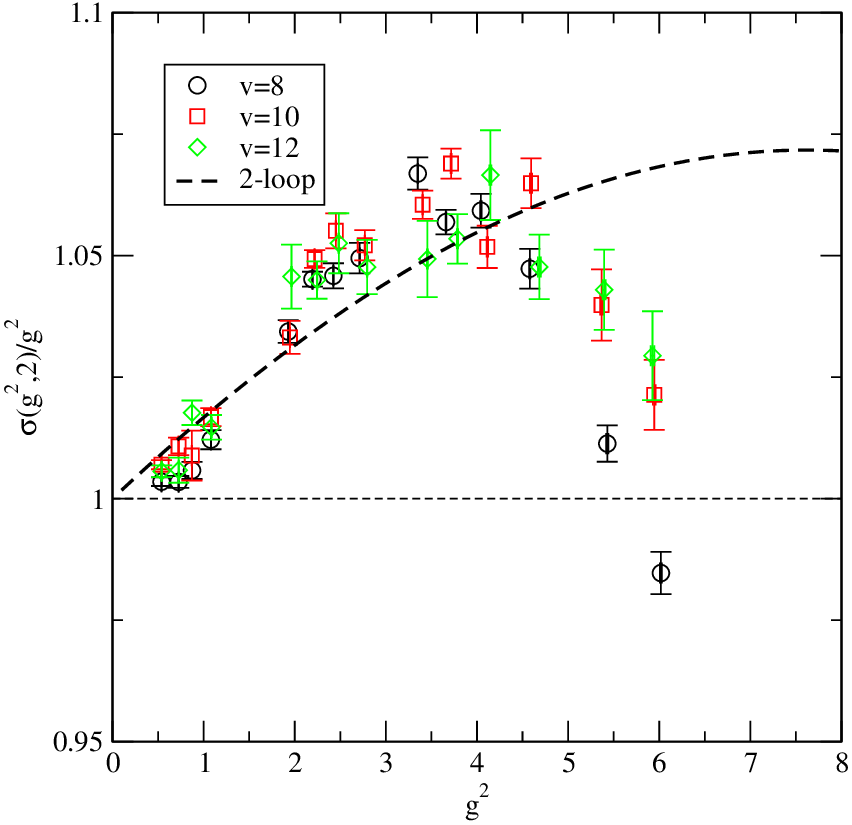} 
  \includegraphics[width=0.45\textwidth]{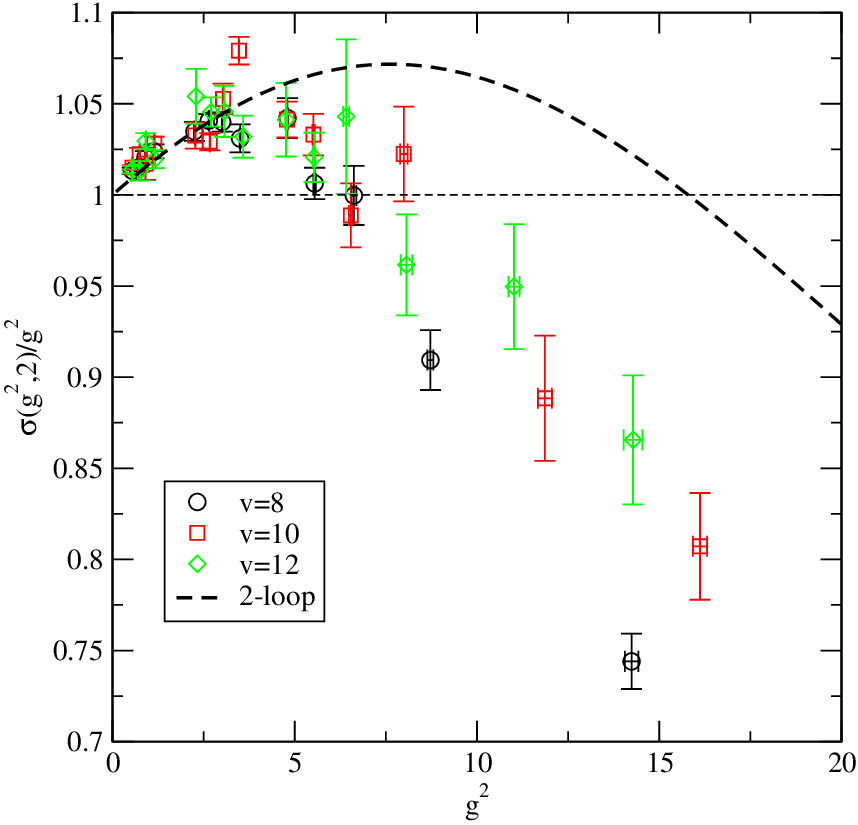}
\caption[b]{
  The lattice step scaling function.
  On the left, $c_t=0.35, \tau_0 = 0.0375\log(1+2 g^2) $.
  On the right, $c_t=0.5, \tau_0 = 0.1\log(1+0.4g^2)$.
}
\label{fig:sigmalat35}
\end{figure}

We show our current results for the lattice step scaling function and the continuum limit (\ref{contfit}) in figures \ref{fig:sigma_lat_cont} and \ref{fig:sigmalat35}.
In each case the running seems to diverge from 2-loop perturbation theory after
 {\kari
 } $g^2 \approx 4$, but 
this is likely to be {\kari at least partly} a discretisation error since the smallest lattice diverges more quickly than the larger ones. The results at $c_t=0.4$ are accurate enough to study the continuum limit. In this case, the step scaling function follows the perturbative value in the entire region, up to $g^2=7$. In figure \ref{fig:sigmalat35}, we study the sensitivity of the results to the precise value of $c_t$ used.

\section{Conclusions}
We have studied the running coupling in the SU(2) lattice gauge theory with 8 fermions in the fundamental representation.
Using the gradient flow coupling and the flow time correction \ref{taucorrection} we have been able to obtain a good continuum limit at small coupling. Even at larger coupling, the results seem to follow perturbation theory and the running remains slow. There is no direct evidence of a fixed point, although one could be expected to lie at a large coupling.

\section{Acknowledgments}

We thank A. Hasenfratz and A. Ramos for helpful discussion.
This work is supported by the Academy of Finland
grants 114371, {\kari 1134018 and 1267286},
the Danish National Research Foundation
DNRF:90 grant and by a Lundbeck Foundation Fellowship grant.
T.K. is also
funded by the Danish Institute for Advanced Study and 
T.R. by the Magnus Ehrnrooth foundation. 
The simulations were performed at the Finnish IT
Center for Science (CSC) in Espoo, Finland,
on the Fermi supercomputer at Cineca in Bologna, Italy
and on the K computer at Riken AICS in Kobe, Japan. 
Parts of the simulation program have been derived from
the MILC lattice simulation program \cite{MILC}.


\begin{thebibliography}{99}

\bibitem{Ohki:2010sr} 
  H.~Ohki, T.~Aoyama, E.~Itou, M.~Kurachi, C.-J.~D.~Lin, H.~Matsufuru, T.~Onogi and E.~Shintani {\it et al.},
  PoS LATTICE {\bf 2010}, 066 (2010)
  [arXiv:1011.0373 [hep-lat]].

\bibitem{Huang:2014xwa} 
  C.~Y.-H.~Huang, C.-J.~D.~Lin, K.~Ogawa, H.~Ohki and E.~Rinaldi,
  arXiv:1410.8698 [hep-lat].

\bibitem{Luscher:1991wu} 
  M.~Luscher, P.~Weisz and U.~Wolff,
  Nucl.\ Phys.\ B {\bf 359}, 221 (1991).

  
\bibitem{Bursa:2010xn}
  F.~Bursa, L.~Del Debbio, L.~Keegan, C.~Pica and T.~Pickup,
  Phys.\ Lett.\ B {\bf 696} (2011) 374
  [arXiv:1007.3067 [hep-ph]].

\bibitem{Karavirta:2011zg}
  T.~Karavirta, J.~Rantaharju, K.~Rummukainen and K.~Tuominen,
  JHEP {\bf 1205} (2012) 003
  [arXiv:1111.4104 [hep-lat]].
  
\bibitem{Hayakawa:2013maa}
  M.~Hayakawa, K.-I.~Ishikawa, S.~Takeda, M.~Tomii and N.~Yamada,
  Phys.\ Rev.\ D {\bf 88} (2013) 9,  094506
  [arXiv:1307.6696 [hep-lat]].

\bibitem{Appelquist:2013pqa}
  T.~Appelquist, R.~C.~Brower, M.~I.~Buchoff, M.~Cheng, G.~T.~Fleming, J.~Kiskis, M.~F.~Lin and E.~T.~Neil {\it et al.},
  Phys.\ Rev.\ Lett.\  {\bf 112} (2014) 111601
  [arXiv:1311.4889 [hep-ph]].


  
\bibitem{Luscher:2011bx} 
  M.~Luscher and P.~Weisz,
  JHEP {\bf 1102}, 051 (2011)
  [arXiv:1101.0963 [hep-th]].

\bibitem{Fritzsch:2013je} 
  P.~Fritzsch and A.~Ramos,
  JHEP {\bf 1310}, 008 (2013)
  [arXiv:1301.4388 [hep-lat]].

\bibitem{Capitani:2006ni}
  S.~Capitani, S.~Durr and C.~Hoelbling,
  JHEP {\bf 0611} (2006) 028
  [hep-lat/0607006].
 
\bibitem{Luscher:2010iy} 
  M.~Luscher,
  JHEP {\bf 1008}, 071 (2010)
  [arXiv:1006.4518 [hep-lat]].
 
\bibitem{Fodor:2012qh} 
  Z.~Fodor, K.~Holland, J.~Kuti, D.~Nogradi and C.~H.~Wong,
  PoS LATTICE {\bf 2012}, 050 (2012)
  [arXiv:1211.3247 [hep-lat]].
   
\bibitem{Luscher:1993gh} 
  M.~Luscher, R.~Sommer, P.~Weisz and U.~Wolff,
  Nucl.\ Phys.\ B {\bf 413}, 481 (1994)
  [hep-lat/9309005].
   
\bibitem{Rantaharju:2013bva} 
  J.~Rantaharju,
  PoS Lattice {\bf 2013}, 084 (2013)
  [arXiv:1311.3719 [hep-lat]].
 
\bibitem{Cheng:2014jba} 
  A.~Cheng, A.~Hasenfratz, Y.~Liu, G.~Petropoulos and D.~Schaich,
  JHEP {\bf 1405}, 137 (2014)
  [arXiv:1404.0984 [hep-lat]].
  
  
  \bibitem{MILC}
  http://physics.utah.edu/$\sim$detar/milc.html 
 





 
\end{thebibliography}
\end{document}